\documentclass[fleqn,usenatbib]{mnras}

\usepackage{newtxtext,newtxmath}

\usepackage[T1]{fontenc}

\DeclareRobustCommand{\VAN}[3]{#2}
\let\VANthebibliography\thebibliography
\def\thebibliography{\DeclareRobustCommand{\VAN}[3]{##3}\VANthebibliography}

\usepackage{multicol}
\usepackage{graphicx}
\usepackage{subcaption}
\usepackage{amsmath}	

\title[Constraints on EGMF from GRB 221009A]{First constraints on the strength of the extragalactic magnetic field \\ from $\gamma$-ray observations of GRB 221009A}

\author[T.A. Dzhatdoev et al.]
{Timur A. Dzhatdoev,$^{1,2}$\thanks{E-mail: timur1606@gmail.com} Egor I. Podlesnyi,$^{3}$ and Grigory I. Rubtsov$^{1}$ \\
$^{1}$ Institute for Nuclear Research of the Russian Academy of Sciences, 60th October Anniversary Prospect 7a, Moscow 117312, Russia \\
$^{2}$ Federal State Budget Educational Institution of Higher Education, M.V. Lomonosov Moscow State University, \\ Skobeltsyn Institute of Nuclear Physics (SINP MSU), 1(2), Leninskie gory, GSP-1, 119991 Moscow, Russia \\
$^{3}$ Norwegian University for Science and Technology (NTNU), Institutt for fysikk, Trondheim, Norway}

\date{Accepted XXX. Received YYY; in original form ZZZ}

\pubyear{2023}

\begin{document}
\label{firstpage}
\pagerange{\pageref{firstpage}--\pageref{lastpage}}
\maketitle

\begin{abstract}
The extragalactic magnetic field (EGMF) could be probed with $\gamma$-ray observations of distant sources. Primary very high energy (VHE) $\gamma$-rays from these sources absorb on extragalactic background light photons, and secondary electrons/positrons from the pair production acts create cascade $\gamma$-rays. These cascade $\gamma$-rays could be detected with space $\gamma$-ray telescopes such as \textit{Fermi}-LAT. The $\gamma$-ray burst GRB 221009A was an exceptionally bright transient well suited for intergalactic $\gamma$-ray propagation studies. Using publicly-available \textit{Fermi}-LAT data, we obtain upper limits on the spectrum of delayed emission from GRB 221009A during the time windows of 10, 30, and 90 days after the burst, and compare these with model spectra calculated for various EGMF strengths $B$, obtaining constraints on $B$. We show that the values of $B$ between $10^{-20}$ G and $10^{-18}$ G are excluded.
\end{abstract}

\begin{keywords}
gamma-ray burst: individual: GRB 221009A --- magnetic fields --- gamma-rays: general --- methods: data analysis --- methods: numerical
\end{keywords}



\section{Introduction}

GRB 221009A, an exceptionally bright \citep{Williams2023,Lesage2023,Burns2023} and relatively nearby (redshift $z = 0.1505$ \citep{deUgartePostigo2022,CastroTirado2022}) $\gamma$-ray burst, has been detected with the WCDA and KM2A arrays of the LHAASO experiment in the energy range $E > 200$ GeV \citep{Cao2023,Huang2022}. In particular, the detection of $\gamma$-rays above the energy of 10 TeV from GRB 221009A was reported.

TeV $\gamma$-rays from GRB 221009A are strongly absorbed on extragalactic background light (EBL) photons by means of the pair production (PP) process ($\gamma \gamma \rightarrow e^{+} e^{-}$). The secondary electrons and positrons\footnote{hereafter collectively called ``electrons''} produced in the PP acts get deflected on the extragalactic magnetic field (EGMF); these secondary electrons then produce cascade $\gamma$-rays by means of the inverse Compton (IC) scattering ($e^{-} \gamma \rightarrow e^{-'} \gamma^{'}$ or $e^{+} \gamma \rightarrow e^{+'} \gamma^{'}$). The energy, angular, and temporal characteristics of this cascade $\gamma$-ray echo are sensitive to the EGMF strength and structure, thus allowing to probe the EGMF with \mbox{$\gamma$-ray} observations of extragalactic sources \citep{Honda1989,Plaga1995,Neronov2007,Neronov2009,Neronov2010,Ichiki2008,Murase2009}. 

Several GRBs were detected in the very high energy (VHE, \mbox{$E > 100$ GeV}) domain before GRB 221009A \citep{MAGIC2019,Abdalla2019,HESS2021,Blanch2020}. For one of them, GRB 190114C, it was shown that the intensity of the cascade $\gamma$-ray echo is below the sensitivity of the operating \mbox{$\gamma$-ray} telescopes even for the EGMF strength $B = 0$ \citep{Dzhatdoev2020} (hereafter D20), and this conclusion was confirmed by \citet{DaVela2023}. \citet{Veres2017} show that for GRB 130427A the cascade echo is detectable with the existing $\gamma$-ray telescopes for $B > 10^{-18}$~G under certain optimistic assumptions on the high intensity of this GRB in the VHE domain. Unfortunately, the GRB 130427A was not detected at TeV energies and thus the latter constraints on $B$ remain conjectural. 

In this {\it Letter}, we report on the constraints on the EGMF strength from $\gamma$-ray observations of GRB 221009A with LHAASO \citep{Ma2022} and \textit{Fermi}-LAT \citep{Atwood2009}. We describe our analysis of \textit{Fermi}-LAT data in Section~\ref{sec:fermi}. Section~\ref{sec:primary} is devoted to the primary $\gamma$-ray spectrum reconstruction. The intergalactic cascade pair echo calculation procedure is outlined in Section~\ref{sec:simulation}. The main results are presented in Section~\ref{sec:results}; then follows a brief discussion (Section~\ref{sec:discussion}) and conclusions (Section~\ref{sec:conclusions}). Appendices contain some details on the data analysis, simulations and results. All graphs in this paper were produced with the ROOT software \citep{Brun1997}.

\section{\textit{Fermi}-LAT data analysis} \label{sec:fermi}

We select \textit{Fermi}-LAT data within 90 days of observation, starting at the \textit{Fermi}-GBM trigger time $T_{0}$ \citep{Lesage2023}. We reconstruct the \textit{Fermi}-LAT spectral energy distribution (SED~=~$E^{2}dN/dE$) of GRB 221009A in the time window from $T_{0}$ to $T_{0} + \delta T_{L}$, where $\delta T_{L} = 2 \times 10^{3}$~s is the duration of the LHAASO observation of the source according to \citet{Cao2023,Huang2022} over which the LHAASO $\gamma$-ray spectrum was measured. The \textit{Fermi}-LAT SED obtained by us is shown in Fig.~\ref{Fig1} as red circles with statistical uncertainties. Some details of this analysis are presented in Appendix~\ref{sec:appendixa}.

\begin{figure}
\centering
\includegraphics[width=8.4cm]{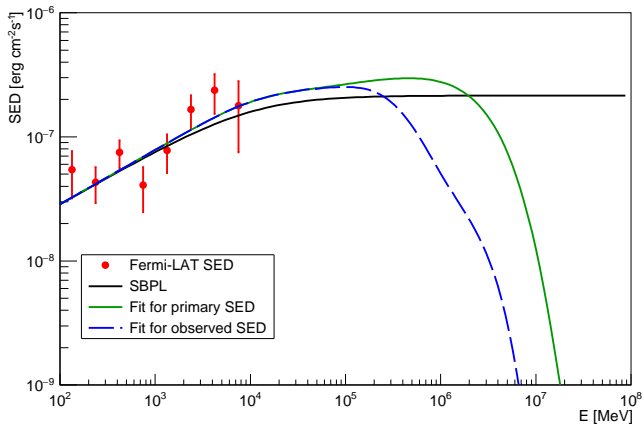}
\caption{\textit{Fermi}-LAT SED of GRB 221009A over the first $2 \times 10^{3}$~s after the \textit{Fermi}-GBM trigger (red circles with statistical uncertainties) together with two options for the possible $\gamma$-ray spectrum of the source in the 100 MeV -- 100 TeV energy range (solid curves; see the text for more details). Blue dashed curve accounts for the effect of $\gamma$-ray absorption on EBL photons vs. green solid curve.}
\label{Fig1}
\end{figure}

We derive upper limits (95 \% C.L.) on the SED of the cascade $\gamma$-ray echo from GRB 221009A, starting at $T_{0} + \delta T_{A}$, where $\delta T_{A} = 2\times10^{5}$~s is an approximate duration of the $\gamma$-ray afterglow of GRB 221009A visible with \textit{Fermi}-LAT \citep{Stern2023}, and ending at $T_{0} + \delta T_{A} + \delta T_{E}$, with three options for $\delta T_{E}$ = 10, 30, and 90 days. The results for the upper limits are shown in Fig.~\ref{Fig2}, \ref{Fig3}, \ref{Fig4}, and \ref{Fig5} (red horizontal bars with downward arrows). Some details of this analysis are presented in Appendix~\ref{sec:appendixa} as well.

\section{The primary \texorpdfstring{$\gamma$}{g}-ray spectrum of GRB 221009A} \label{sec:primary}

\subsection{Pre-publication estimates}

Even before the publication of \citet{Cao2023}, it was possible to obtain some estimates of the primary $\gamma$-ray spectrum for GRB 221009A. We fit the \textit{Fermi}-LAT spectrum shown in Fig.~\ref{Fig1} with a power-law function, obtaining the best-fit index $\gamma_{1} = 1.56$. The broadband $\gamma$-ray SED of GRB 221009A could be characterised with the following smoothly broken power law (SBPL) function:
\begin{equation}
E^{2}\frac{dN}{dE} = K_{s}  \left( \frac{E}{E_{s}} \right)^{2} \left( \frac{E}{E_{s}} \right)^{-\gamma_{1}} \left[1+ \left( \frac{E}{E_{b}} \right)^{\epsilon} \right]^{-(\gamma_{2}-\gamma_{1})/\epsilon}, \label{eq1}
\end{equation}
where $K_{s} = 5.38 \times 10^{-8}$ erg$\:$cm$^{-2}$s$^{-1}$ is the normalization factor, $\gamma_{2} = 2$, $E_{b}= 10$ GeV, $\epsilon = 1$, and $E_{s} = 422$ MeV is the reference energy. The corresponding SED is shown in Fig.~\ref{Fig1} as solid black curve. Two additional options for the primary $\gamma$-ray SED, namely, the primary power law (PWL) and log-parabolic (LP) SEDs, are covered in Appendix~\ref{sec:appendixb}.

\subsection{Post-publication estimates}

After the release of the LHAASO-WCDA dataset on GRB 221009A \citet{Cao2023}, it became possible to constrain the primary $\gamma$-ray spectrum of GRB 221009A more tightly than before the publication of this paper. Here we use the same approach as in D20 except that \citet{Cao2023} presented five partial spectra over different time intervals, and not one spectrum as was the case for GRB 190114C. After fitting all five partial SEDs with the $\propto E^{-\gamma}exp(-E/E_{c})$ functional form and obtaining tables for the corresponding $F_{1}(E), F_{2}(E), (...), F_{5}(E)$ functions, the average primary $\gamma$-ray SED over 2000~s of observation is computed:
\begin{equation}
E^{2}\frac{dN}{dE} = \frac{\delta t_{1}F_{1}(E) + \delta t_{2}F_{2}(E) + (...) + \delta t_{5}F_{5}(E)}{2000 \: s}, \label{eq2}
\end{equation}
where $\delta t_{n}$ is the duration of the $n$th observation episode, $n=1,2,$...,$5$.

Finally, we utilized the LHAASO-KM2A dataset on GRB 221009A \citep{Wang2023} to correct the SED $F_{4}(E)$ for the observation period of 326-900~s. Some details of this procedure are presented in Appendix~\ref{sec:appendixc}.

The result of this procedure --- the time-averaged primary SED (i.e. the SED before the absorption on EBL photons) in the energy range $E > 100 GeV$ --- is shown in Fig.~\ref{Fig1} as green solid curve. The same SED, but after the EBL absorption effect was applied, is shown in the same Figure as long-dashed blue curve. An approximation of the \textit{Fermi}-LAT SED below 10 GeV is shown for consistency.\footnote{this low-energy part of the SED was obtained as an SBPL lower-energy extension of the LHAASO dataset fit}

\section{Simulation of the pair echo from GRB 221009A} \label{sec:simulation}

We calculate the observable SED of the intergalactic cascade pair echo using the publicly available code ELMAG~3.03 \citep{Blytt2020,Kachelriess2012} in the time window from $T_{0} + \delta T_{A}$ to $T_{0} + \delta T_{A} + \delta T_{E}$. The maximal energy of the primary $\gamma$-rays is set to 100 TeV. The general scheme of calculations follows D20.

We assume the fiducial EBL model of \citet{Gilmore2012}. As in D20, the EGMF was modeled as isotropic random nonhelical turbulent field with a Kolmogorov spectrum and Gaussian variance $B_{\mathrm{RMS}}$ (hereafter simply $B$) following the approach of \citet{Giacalone1994,Giacalone1999}. The minimal and maximal EGMF spatial scales were set as $L_{\mathrm{min}} = 5 \times 10^{-4}$~Mpc and $L_{\mathrm{max}} = 5$~Mpc\footnote{this corresponds to the coherence length of $L_{c} \approx 1$~Mpc}, respectively, with 200 field modes in total. Full three-dimensional simulation was employed. The jet opening angle $\theta_{\mathrm{jet}}$ was set to 1 degree, not far from the value obtained in \citet{Cao2023}; however, the conclusions of this paper are almost independent on the value of $\theta_{\mathrm{jet}}$ (see Discussion below). We neglect collective (plasma) energy losses for cascade electrons \citep{Broderick2012,Schlickeiser2012,Schlickeiser2013,Miniati2013,Chang2014,Sironi2014,Menzler2015,Kempf2016,Vafin2018,Vafin2019,Perry2021}. For sufficiently large values of $B$, the width of the pair echo's observable angular distribution $\theta_{\mathrm{obs}}$ is comparable to or larger than the width of the point spread function (PSF) of \textit{Fermi}-LAT $\theta_{\mathrm{PSF}}$. This could affect the reconstructed point-like spectrum of the source. In what follows we neglect the latter effect since $\theta_{\mathrm{obs}} \ll \theta_{\mathrm{PSF}}$ for the values of $B \lesssim 10^{-18}$ G \citep{Neronov2009}.

\section{Results} \label{sec:results}

\subsection{Pre-publication estimates}

\begin{figure}
\centering
\includegraphics[width=8.4cm]{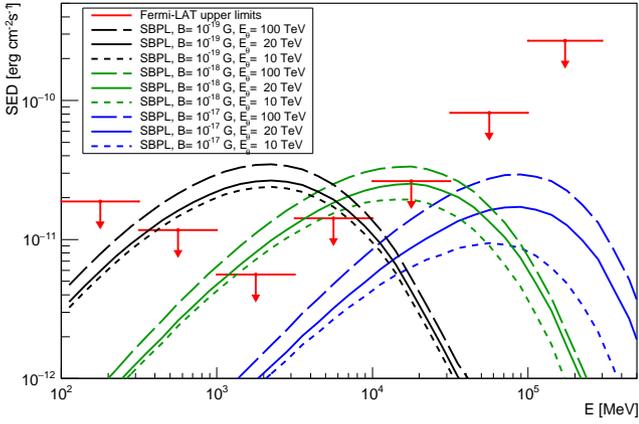}
\caption{\textit{Fermi}-LAT upper limits on the SED of GRB 221009A in the time interval 2$\times10^{5}$~s -- 30~days + 2$\times10^{5}$~s after the \textit{Fermi}-GBM trigger (red horizontal bars with downward arrows) together with model SEDs for various values of $B$ (curves; see the text for more details).}
\label{Fig2}
\end{figure}

Here we report our results for the case of the SBPL primary spectrum option (solid black curve in Fig.~\ref{Fig1}) and $\delta T_{E} = 30$ days. The simulated $\gamma$-ray spectra of the cascade echo are shown in Fig.~\ref{Fig2} for $B = 10^{-19}$ G (black curves), $B = 10^{-18}$ G (green curves) and $B = 10^{-17}$ G (blue curves). An additional step-function cutoff in the primary $\gamma$-ray spectrum was introduced; solid curves correspond to the cutoff energy $E_{\theta} = 20$~TeV, short-dashed curves --- to $E_{\theta} = 10$~TeV, long-dashed curves --- to $E_{\theta} = 100$~TeV. We conclude that the case of $B = 10^{-18}$ G is excluded even for $E_{\theta} = 10$~TeV. Results for the case of the PWL and LP SEDs are presented in Appendix~\ref{sec:appendixb} (See Fig.~\ref{FigB2} and Fig.~\ref{FigB3}, respectively).

\subsection{Post-publication estimates}

\begin{figure}
\centering
\includegraphics[width=8.4cm]{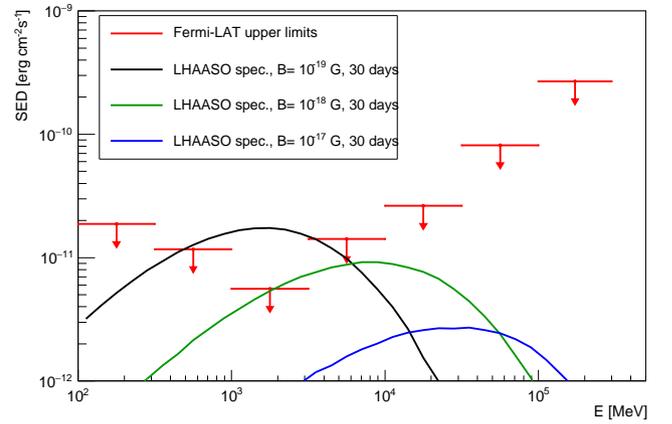}
\caption{Same as in Fig.~\ref{Fig2}, but for the case of the post-publication primary SED estimate.}
\label{Fig3}
\end{figure}

\begin{figure}
\centering
\includegraphics[width=8.4cm]{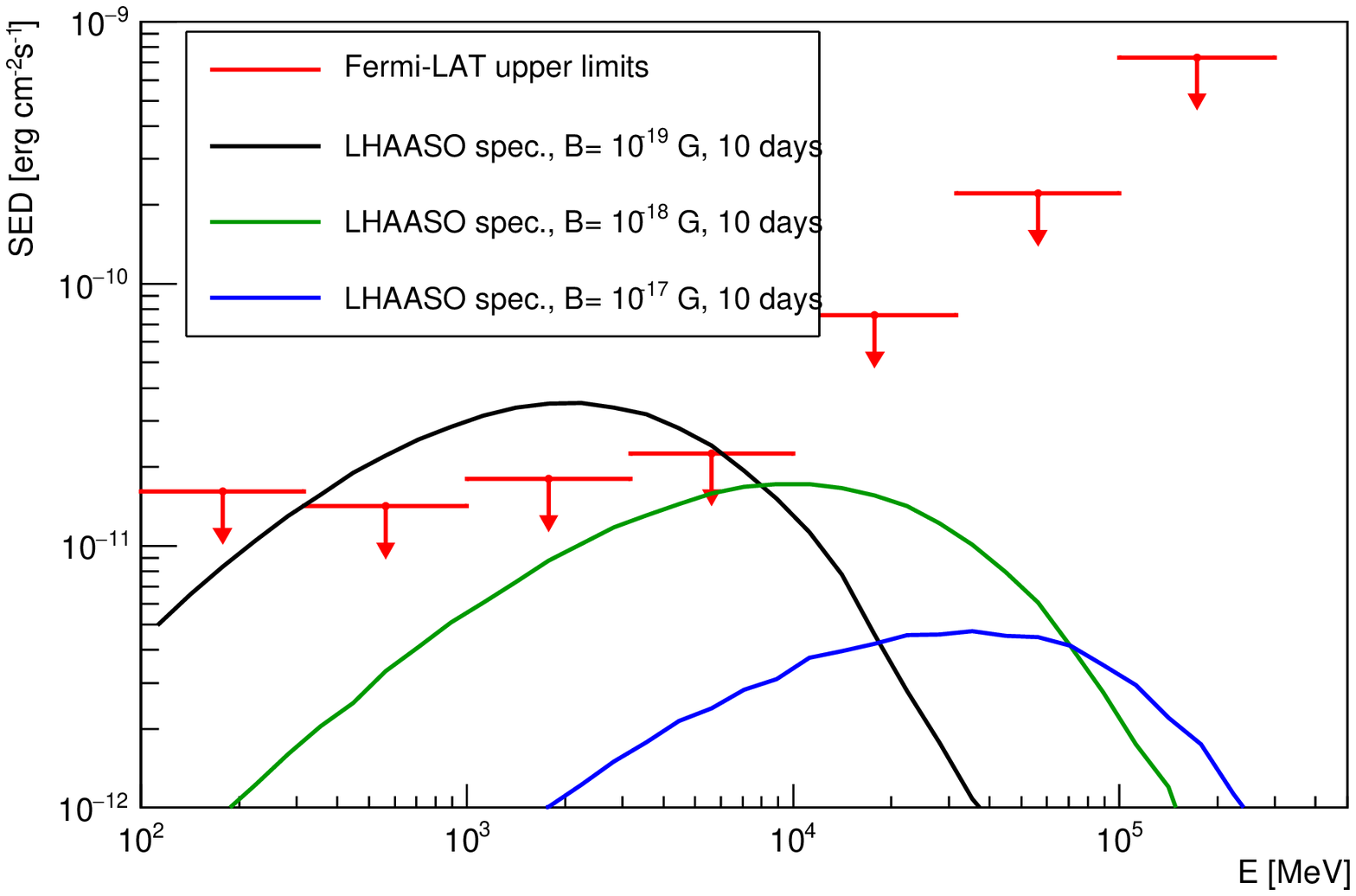}
\caption{Same as in Fig.~\ref{Fig3}, but for the case of $\delta T_E = 10$~days.}
\label{Fig4}
\end{figure}

\begin{figure}
\centering
\includegraphics[width=8.4cm]{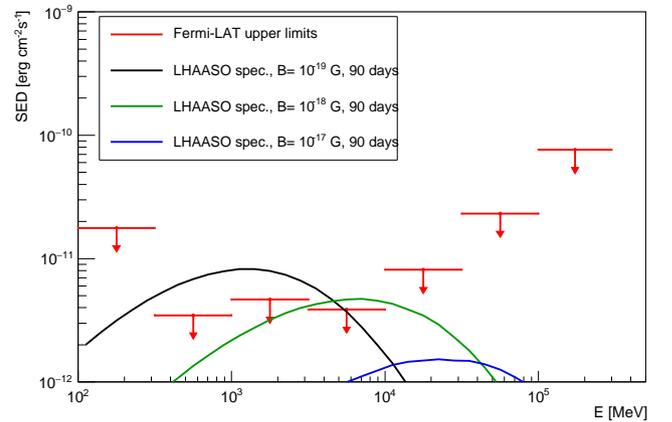}
\caption{Same as in Fig.~\ref{Fig3}, but for the case of $\delta T_E = 90$~days.}
\label{Fig5}
\end{figure}

Here we report our results for the case of the post-publication estimate of the primary $\gamma$-ray SED (solid green curve in Fig.~\ref{Fig1}). Fig.~\ref{Fig3} shows the results for $\delta T_{E} = 30$ days, Fig.~\ref{Fig4} --- for $\delta T_{E} = 10$ days, Fig.~\ref{Fig5} --- for $\delta T_{E} = 90$ days.

A plot including additional EGMF strength values in range from $B = 10^{-21}$~G to $B = 3 \times 10^{-17}$~G for the case of $\delta T_{E} = 90$ days is presented in Appendix~\ref{sec:appendixd} (see Fig.~\ref{FigD1}), leading to the exclusion of the range of $B$ values from $B = 10^{-20}$~G to $B = 10^{-18}$~G. The option of $B < 10^{-20}$~G is already excluded as stems from the negative results of the search for cascade echo from blazars \citep{Dermer2011,Taylor2011,Finke2015,Podlesnyi2022}.

Finally, the comparison of cascade echo SEDs for the case of $\delta T_{E} = 90$ days and two different values of $\delta T_{A} = 2 \times 10^{5}$~s and $\delta T_{A} = 0$ is presented in Appendix~\ref{sec:appendixe}. For $\delta T_{A} = 0$, the SED curves for $B = 10^{-18}$~G and $B = 10^{-17}$~G successively branch down from the SED curve for $B = 10^{-19}$~G. This behaviour of the SEDs was reported in D20 (see discussion of Fig.~3 in D20).

\section{Discussion} \label{sec:discussion}

The obtained results are directly relevant for a large-scale EGMF with the coherence length $\lambda > 10-100$ kpc. In this case the typical electron energy loss length $L_{E-e} < \lambda$ \citep{Neronov2009}. In the opposite case of a ``turbulent'' EGMF ($L_{E-e} > \lambda$) the resulting limits on $B$ become even stronger \citep{Neronov2009}.

The obtained constraints depend on the assumed EBL model. Similar to D20, we performed calculations of the pair echo spectrum for a modified EBL model with the intensity normalization factor $K_{\mathrm{EBL}} = 0.7$. The resulting borderline values of $B$ typically change only slightly, within $\approx 20$ \%. 

\subsection{Dependence of results on the jet profile}

The obtained results may also depend on the exact shape of the GRB 221009A jet profile, the opening angle of the jet $\theta_{\mathrm{jet}}$ and the value of the angle $\theta_{\mathrm{view}}$ at which the observer looks into the jet. The typical deflection angle of cascade electrons \citep{Neronov2009}; \citep[see eq. (4)]{Dzhatdoev2020} for $B \lesssim 10^{-18}$~G $\theta_{B} \ll \theta_{\mathrm{jet}}$, thus the dependence of our results on $\theta_{\mathrm{jet}}$ is small.\footnote{\citet{Cao2023} estimate $\theta_{\mathrm{jet}} \approx 0.6$ degree}

\subsection{Comparison of results with other authors}

After the first version of our paper was sent to arXiv and to the journal, two other papers on the EGMF constraints from GRB 221009A appeared \citep{Huang2023,Vovk2023}. The approach of \citet{Huang2023} is similar to the one of D20 and the present work; however, they assumed a slightly different EBL model of \citet{SaldanaLopez2021}. \citet{Huang2023} find that the values of $B \le 10^{-18.5}$ G are excluded. The difference of EBL models and time windows assumed may account for the slightly different results of \citet{Huang2023} compared to the ones obtained in the present work.

\citet{Vovk2023}, on the other hand, utilize the \textit{Fermi}-LAT light curve of GRB 221009A in the time range of $10^{-3} - 10$ days to obtain constraints on $B$. They exclude $B \le 10^{-19}$ G. Given the different approach of \citet{Vovk2023} compared to our work, such a difference of the results could be expected.

\subsection{Observational prospects}

We note that the advent of the next-generation space $\gamma$-ray telescopes such as MAST \citep{Dzhatdoev2019} could dramatically improve the pair echo detectability prospects. The improved sensitivity of the Cherenkov Telescope Array (CTA) \citep{Actis2011,Acharya2013} in the energy range of 100 GeV -- 10 TeV could significantly facilitate the measurement of the intrinsic spectrum, reducing the uncertainty of the pair echo characteristics.

\section{Conclusions} \label{sec:conclusions}

Using the $\gamma$-ray observations of GRB 221009A with LHAASO and \textit{Fermi}-LAT, we were able, for the first time, to obtain constraints on the EGMF strength from GRB emission. We show that the values of $B \leq 10^{-18}$ G are excluded for all values of the EGMF coherence length and for all considered models of the primary $\gamma$-ray spectrum.

\section*{Acknowledgements}

The work of TD and GR was supported by the Russian Science Foundation, grant no. 22-12-00253.

\section*{Data Availability}

The datasets used for the \textit{Fermi}-LAT data analysis presented in this work are publicly available at the \textit{Fermi}-LAT data server\footnote{\url{https://fermi.gsfc.nasa.gov/cgi-bin/ssc/LAT/LATDataQuery.cgi}}. The ELMAG 3.03 code is provided by its authors\footnote{\url{https://elmag.sourceforge.net}}.

\bibliographystyle{mnras}
\bibliography{GRB221009A-EGMF} 

\appendix
\section{\textit{Fermi}-LAT data analysis details} \label{sec:appendixa}

The region of interest (ROI) is a circle with the radius of 20$^{\circ}$, centred at the position of the GRB ($\alpha_{J2000} = 288.264 ^{\circ}$, $\delta_{J2000} = 19.773 ^{\circ}$). We have applied the energy selection from 100 MeV to 300 GeV. For other selection parameters, we use standard recommendations for point-like sources\footnote{\url{https://fermi.gsfc.nasa.gov/ssc/data/analysis/documentation/Cicerone/Cicerone_Data_Exploration/Data_preparation.html}}. 

We then perform unbinned likelihood analysis of the selected data with FermiTools version 2.20\footnote{\url{https://fermi.gsfc.nasa.gov/ssc/data/analysis/documentation/}} assuming \texttt{P8R3\_SOURCE\_V3} instrument response functions. We construct a model of observed emission including the following sources: 1)~GRB 221009A itself, modeled as a pointlike source with power-law spectrum at the center of the ROI, 2) all sources from the Fermi 8-Year Point Source Catalog (4FGL) \citep{Abdollahi2020} located within 17$^{\circ}$ from the center of the ROI, and 3) galactic and isotropic diffuse $\gamma$-ray backgrounds using models \texttt{gll\_iem\_v07} and \texttt{iso\_P8R3\_SOURCE\_V3\_v1} provided by the \textit{Fermi}-LAT Collaboration. For GRB 221009A, we set both spectral index and normalization as free parameters; normalizations for the diffuse backgrounds are left free.

We first perform the fit in 100 MeV -- 300 GeV energy range keeping free all parameters for the pointlike and extended sources from the 4FGL catalog within 5$^{\circ}$ from the center of the ROI. For the sources between 5$^\circ$ to 17$^\circ$ from the center of the ROI we fix all the parameters at their values from the 4FGL catalog. For the case of the SED measurement over the first $2 \times 10^{3}$~s after the trigger, this procedure is performed over the time window of $10^{5}$ s after the trigger to better constrain the parameters of the steady sources; for the case of the delayed emission search, the relevant time window is from $T_{0} + \delta T_{A}$ to $T_{0} + \delta T_{A} + \delta T_{E}$. In the latter case the energy range is 100 MeV -- 500 GeV instead of 100 MeV -- 300 GeV.

To obtain the GRB 221009A SED (or derive upper limits on it) we repeat the fit in every energy bin of the interest keeping only parameters of GRB 221009A free and fixing all parameters of pointlike and extended 4FGL sources as well as the normalizations of galactic and extragalactic backgrounds. Using this model of the observed emission, we obtain the SED of GRB 221009A over the first $2 \times 10^{3}$~s after the trigger presented in Fig.~\ref{Fig1}. As for the emission from GRB 221009A after $T_{0} + \delta T_{A}$, no significant $\gamma$-ray flux was detected by \textit{Fermi}-LAT, therefore we place upper limits (95 \% C.L.) on the SED of the delayed emission presented in Figs.~\ref{Fig2}--\ref{Fig3}, Fig.~\ref{Fig4}, Fig.~\ref{Fig5} for the values of $\delta T_{E}$ = 30, 10 and 90 days respectively. We follow the procedure similar to the one implemented in the user-contributed \textit{Python} script \texttt{SED.py}\footnote{\url{https://fermi.gsfc.nasa.gov/ssc/data/analysis/user/SED_scripts_v13.1.tgz}} to calculate the presented upper limits.

\section{The case of LP and PWL spectra (pre-publication estimates)} \label{sec:appendixb}

\begin{figure}
\centering
\includegraphics[width=7.4cm]{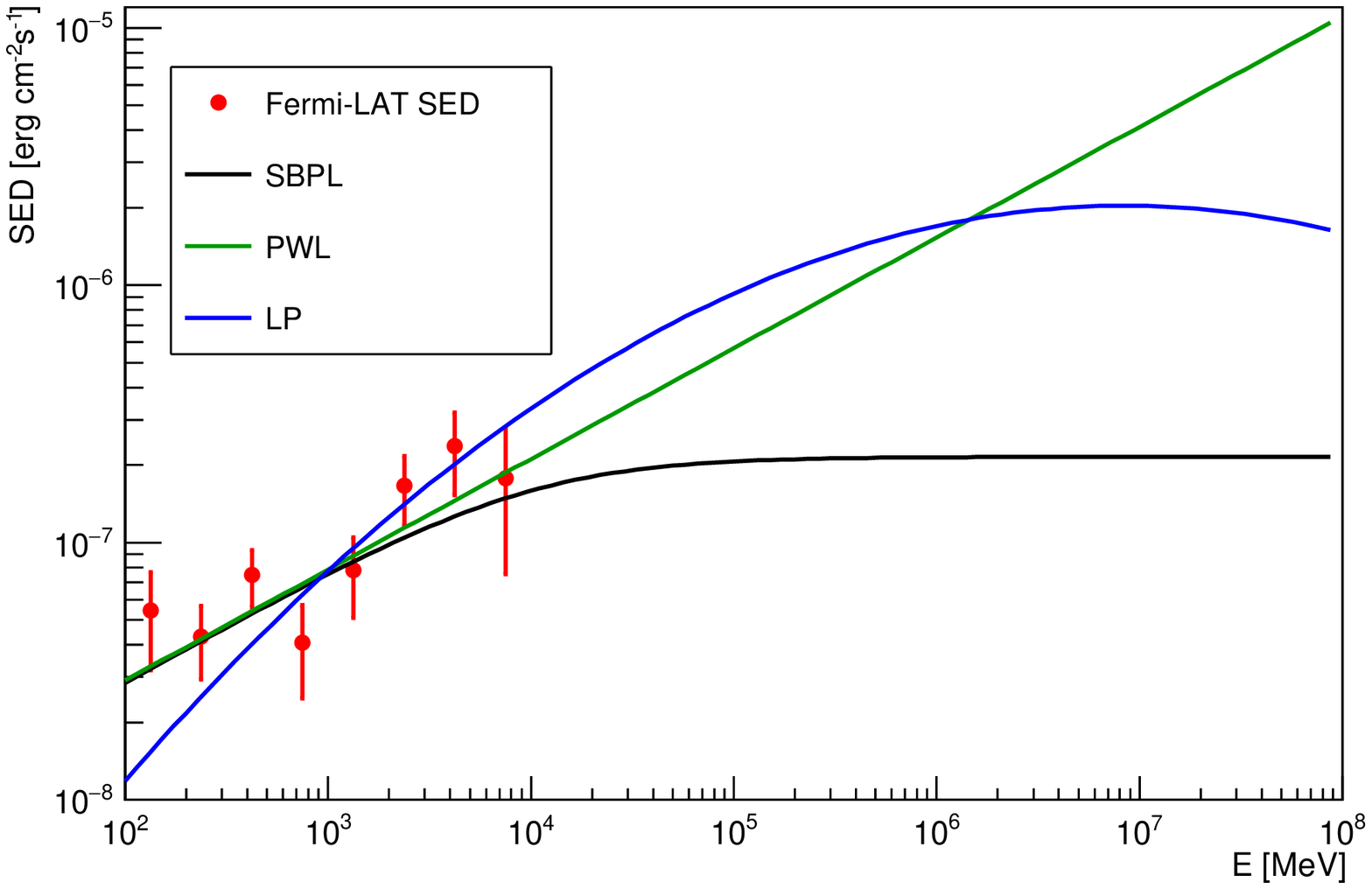}
\caption{\textit{Fermi}-LAT SED of GRB 221009A over the first $2 \times 10^{3}$~s after the \textit{Fermi}-GBM trigger (red circles with statistical uncertainties) together with three options for the possible $\gamma$-ray spectrum of the source in the 100 MeV -- 100 TeV energy range (curves; see the text for more details).}
\label{FigB1}
\end{figure}

\citet{Huang2022} reported the observation of more than $N_{\gamma} = 5\times10^{3}$ $\gamma$-rays from GRB 221009A in the energy range between 500 GeV and 18 TeV. Assuming $N_{\gamma} = 5\times10^{3}$, $\beta \ge 0$ and taking the effective areas of the WCDA and KM2A arrays according to the Supplementary Information for \citet{LHAASO2021}, we estimate the parameters of the LP SED as follows:
\begin{equation}
E^{2}\frac{dN}{dE} = K_{l} \left( \frac{E}{E_{l}} \right)^{2} \left( \frac{E}{E_{l}} \right)^{-\alpha-\beta \ln(E/E_{l})}, \label{eq2}
\end{equation}
where $K_{l} = 8.85 \times 10^{-8}$ erg$\:$cm$^{-2}$s$^{-1}$, $\alpha = 1.57$, $\beta = 0$, and the reference energy $E_{l} = 1.33$ GeV. In this case the best fit is a pure power-law function\footnote{a particular case of the LP function} shown in Fig.~\ref{FigB1} as green line and denoted as PWL. Finally, we consider another option of the primary spectrum (denoted as LP) with $K_{l} = 9.50 \times 10^{-8}$ erg$\:$cm$^{-2}$s$^{-1}$, $\alpha = 1.30$, $\beta = 4 \times 10^{-2}$, and $E_{l} = 1.33$ GeV (shown in Fig.~\ref{FigB1} as blue curve). While performing the fitting for both the PWL and LP cases we added the following additional constraint on the fit: $N_{\gamma} - \delta N_{\gamma} < N_{\gamma-\mathrm{fit}} < N_{\gamma} + \delta N_{\gamma}$ with $\delta N_{\gamma} = \sqrt{N_{\gamma}}$, where $N_{\gamma-\mathrm{fit}}$ is the estimated number of $\gamma$-ray events with the energy $E > 500$ GeV that would be registered with the LHAASO detector.  

\begin{figure}
\centering
\includegraphics[width=7.4cm]{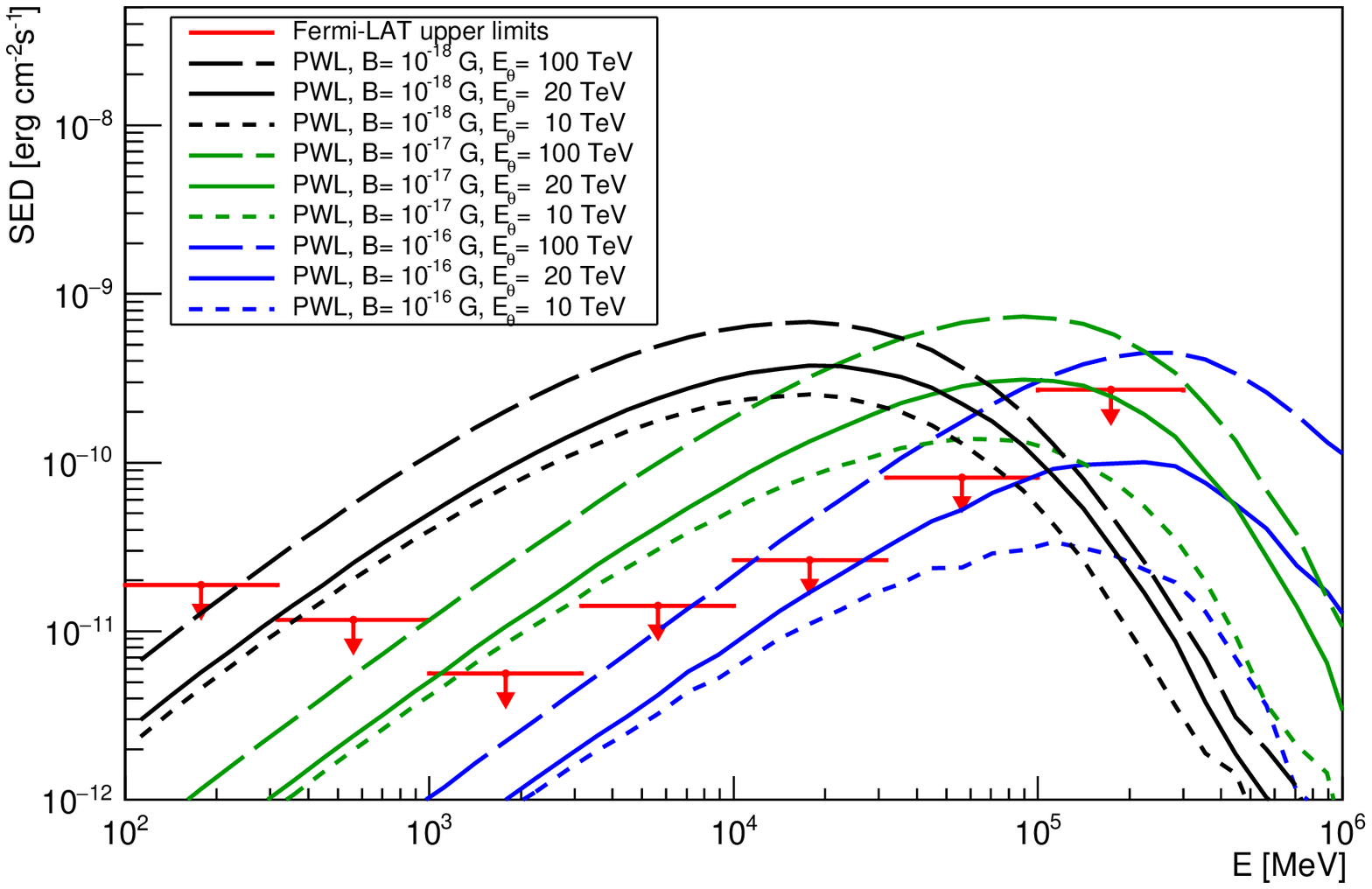}
\caption{Same as in Fig.~\ref{Fig2}, but for the case of the primary power-law spectrum.}
\label{FigB2}
\end{figure}

\begin{figure}
\centering
\includegraphics[width=7.4cm]{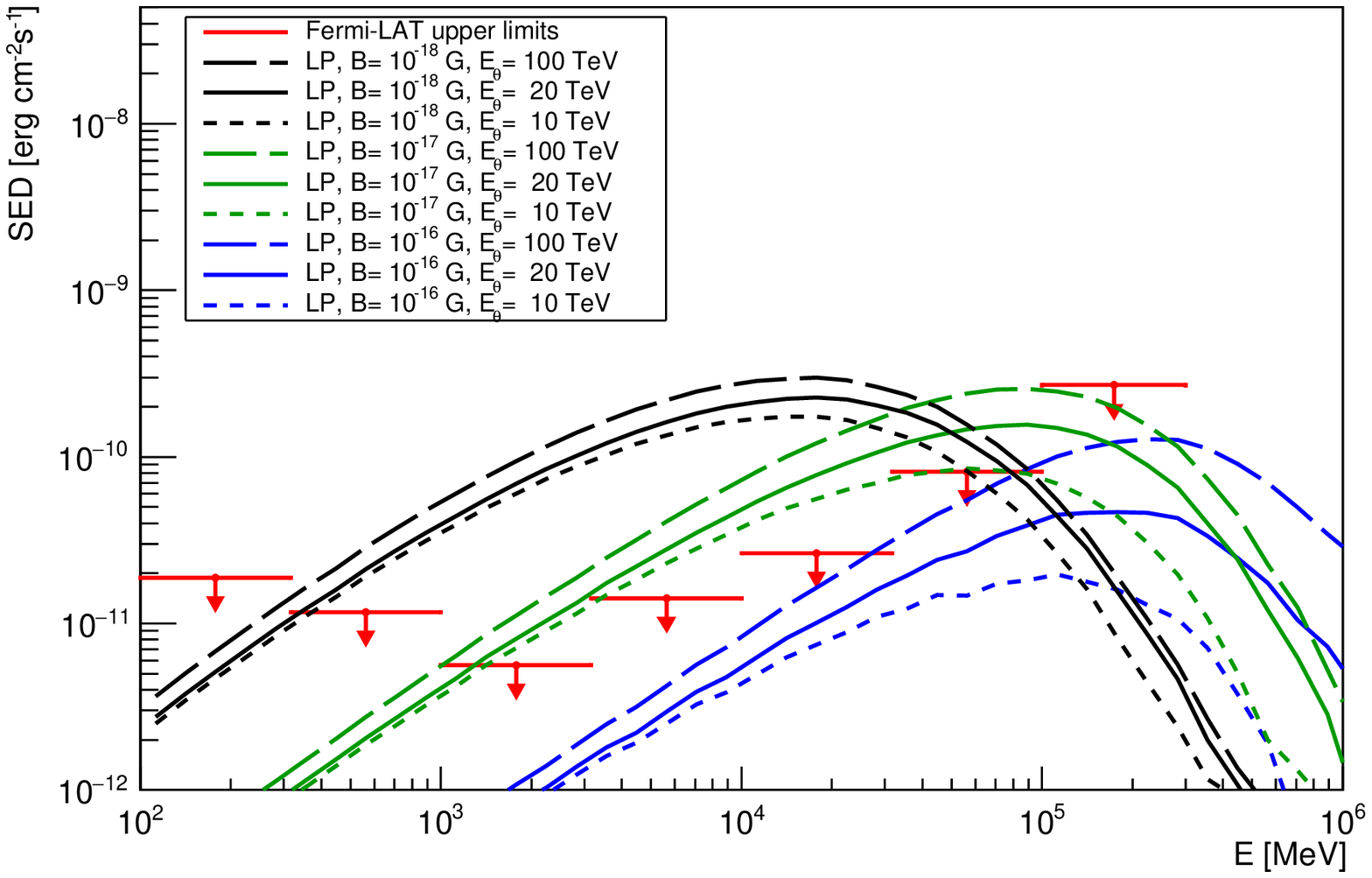}
\caption{Same as in Fig.~\ref{Fig2}, but for the case of the primary log-parabolic spectrum.}
\label{FigB3}
\end{figure}

Results similar to those presented in Fig.~\ref{Fig2} are shown in Fig.~\ref{FigB2} and Fig.~\ref{FigB3} for the case of the power-law and log-parabolic primary spectra (green line and blue curve in Fig.~\ref{FigB1}, respectively). In these cases, the values of (approximately) $B \leq 10^{-16}-10^{-17}$ G could be excluded depending on the value of $E_{\theta}$.

\section{Reconstruction of the primary SED from the LHAASO dataset} \label{sec:appendixc}

The partial spectra of GRB 221009A for five time intervals measured with the LHAASO-WCDA detector were taken from Supplementary Information of \citet{Cao2023} (see their Figure S4C). The fitting of the partial spectra was done in the same manner as in D20 for the case of GRB 190114C.

\begin{figure*}
        \centering
        \begin{multicols}{2}
            \subcaptionbox{$231-240$~s \label{FigC1}}{\includegraphics[width=7.4cm]{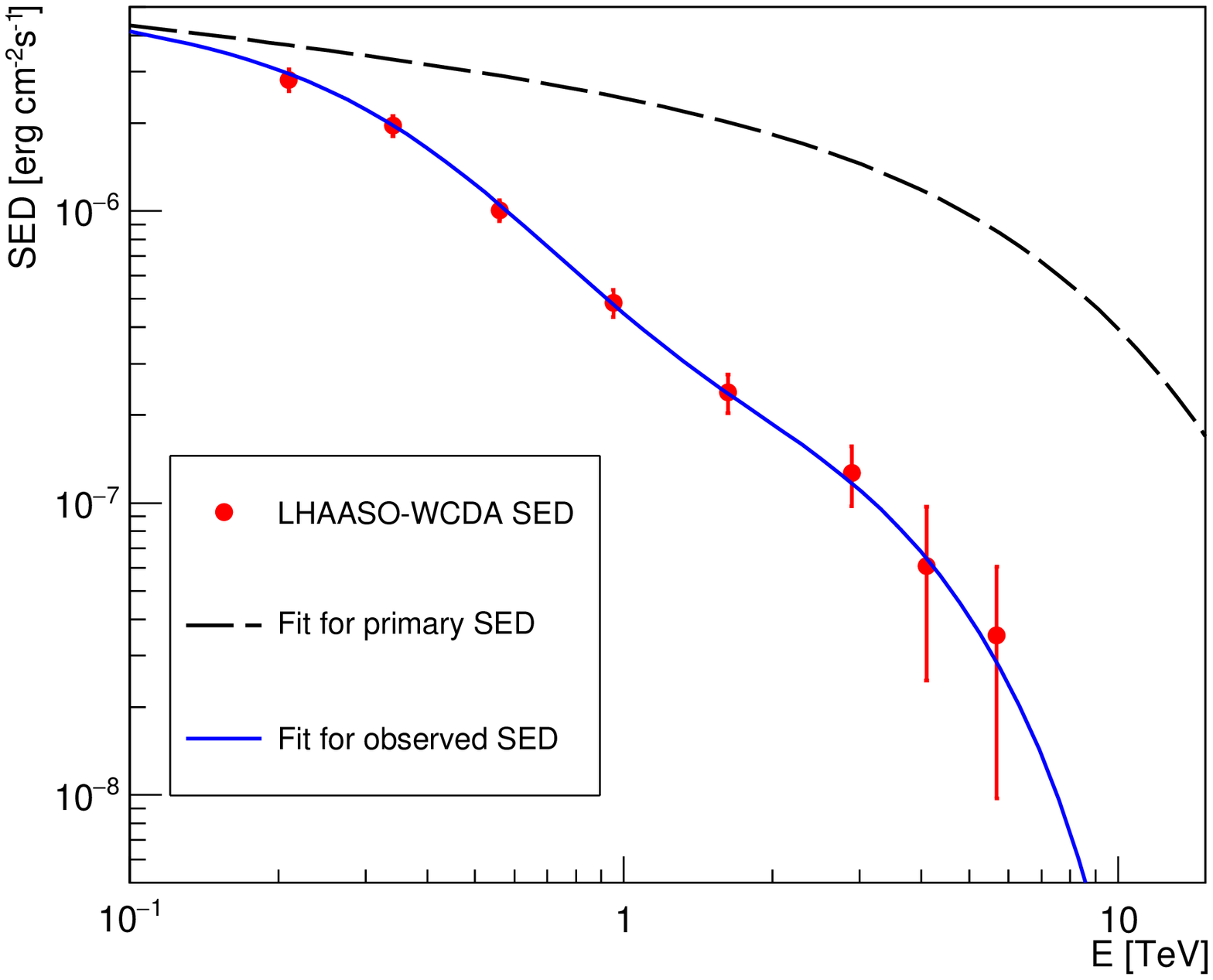}}\par 
            \subcaptionbox{$240-248$~s \label{FigC2}}{\includegraphics[width=7.4cm]{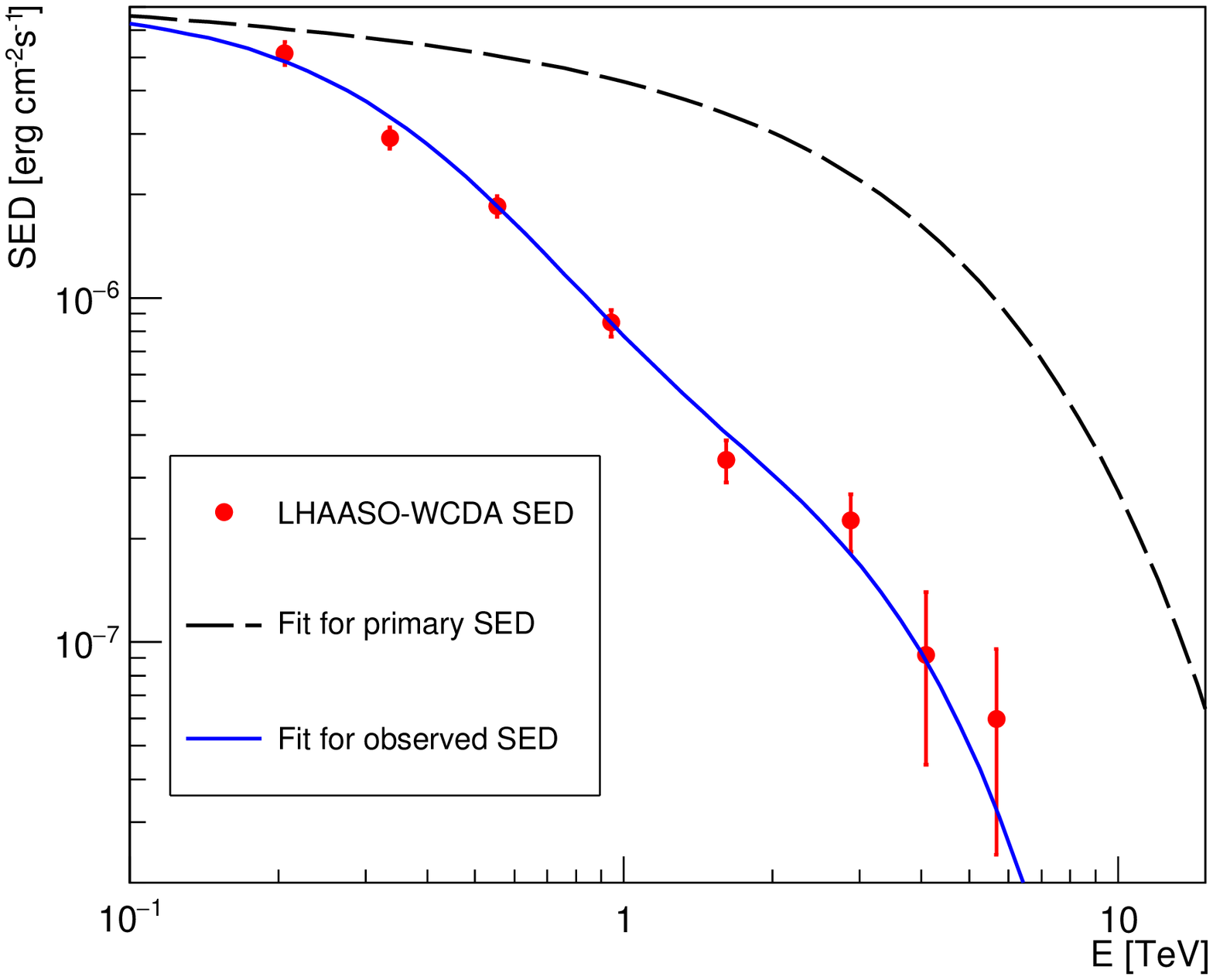}}\par 
        \end{multicols}
        \begin{multicols}{2}
            \subcaptionbox{$248-326$~s \label{FigC3}}{\includegraphics[width=7.4cm]{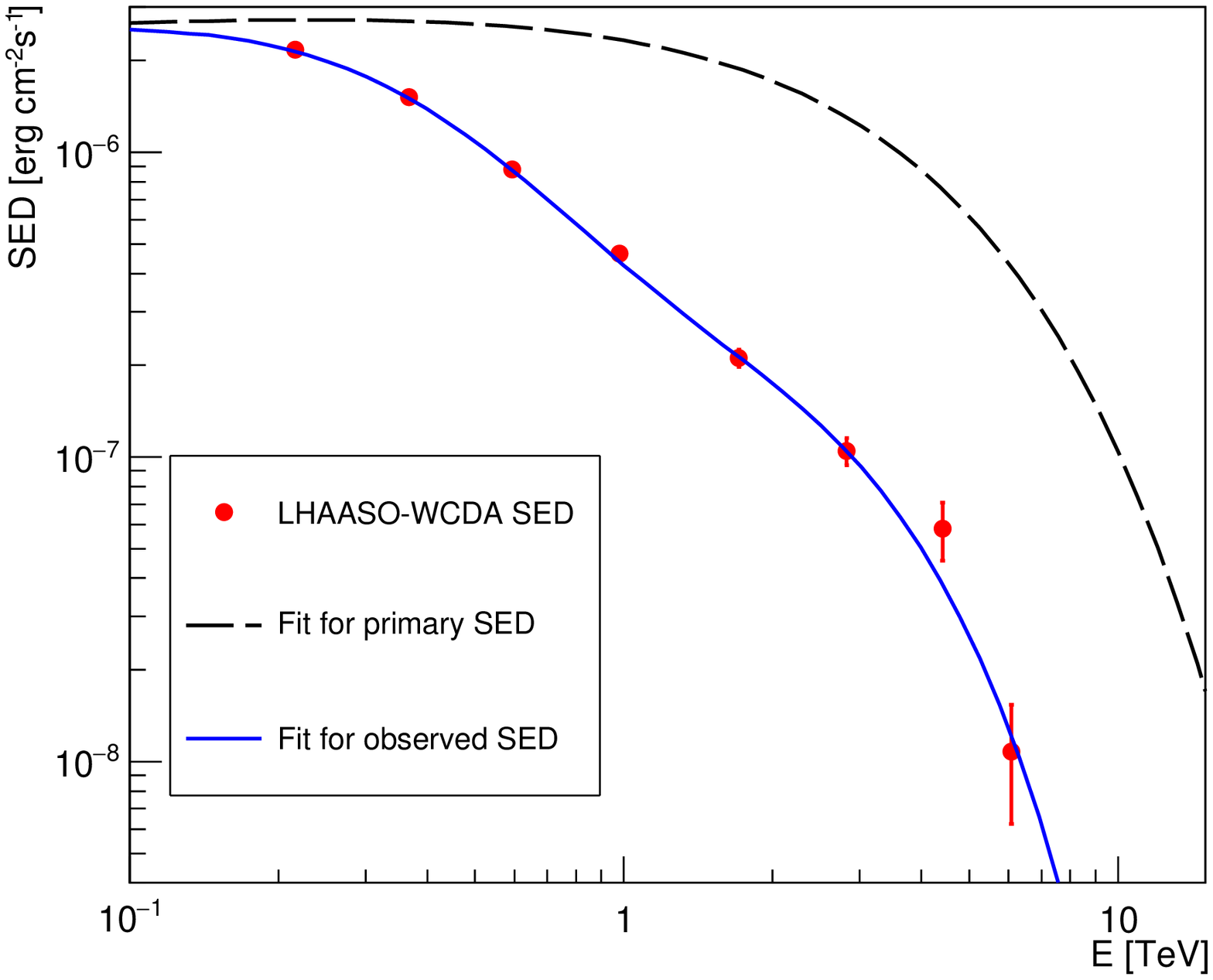}}\par
            \subcaptionbox{Circles: $326-900$~s; squares: $300-900$~s re-scaled to $326-900$~s \label{FigC4}}{\includegraphics[width=7.4cm]{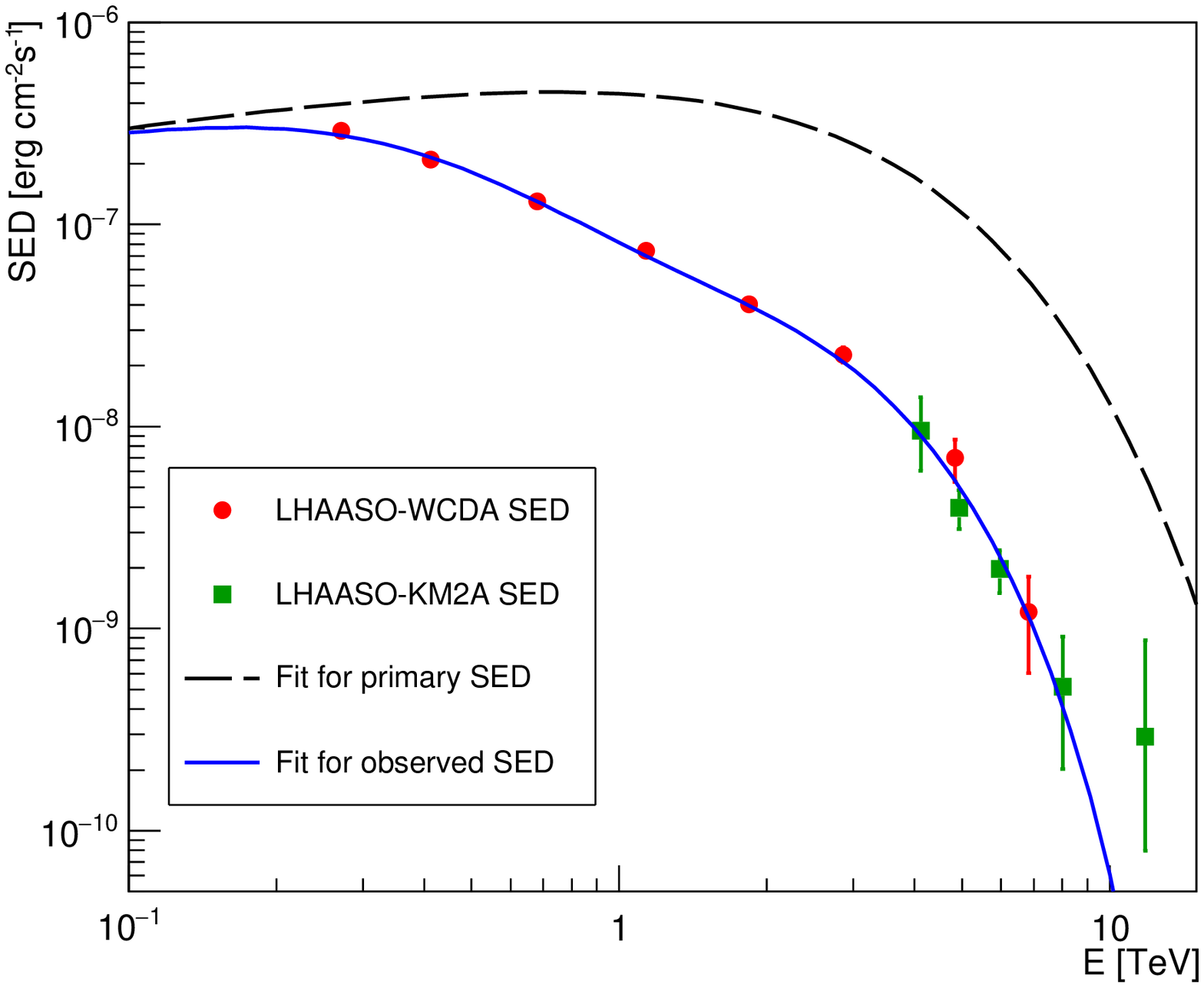}}\par
        \end{multicols}
        \begin{multicols}{2}
            \subcaptionbox{$900-2000$~s \label{FigC5}}{\includegraphics[width=7.4cm]{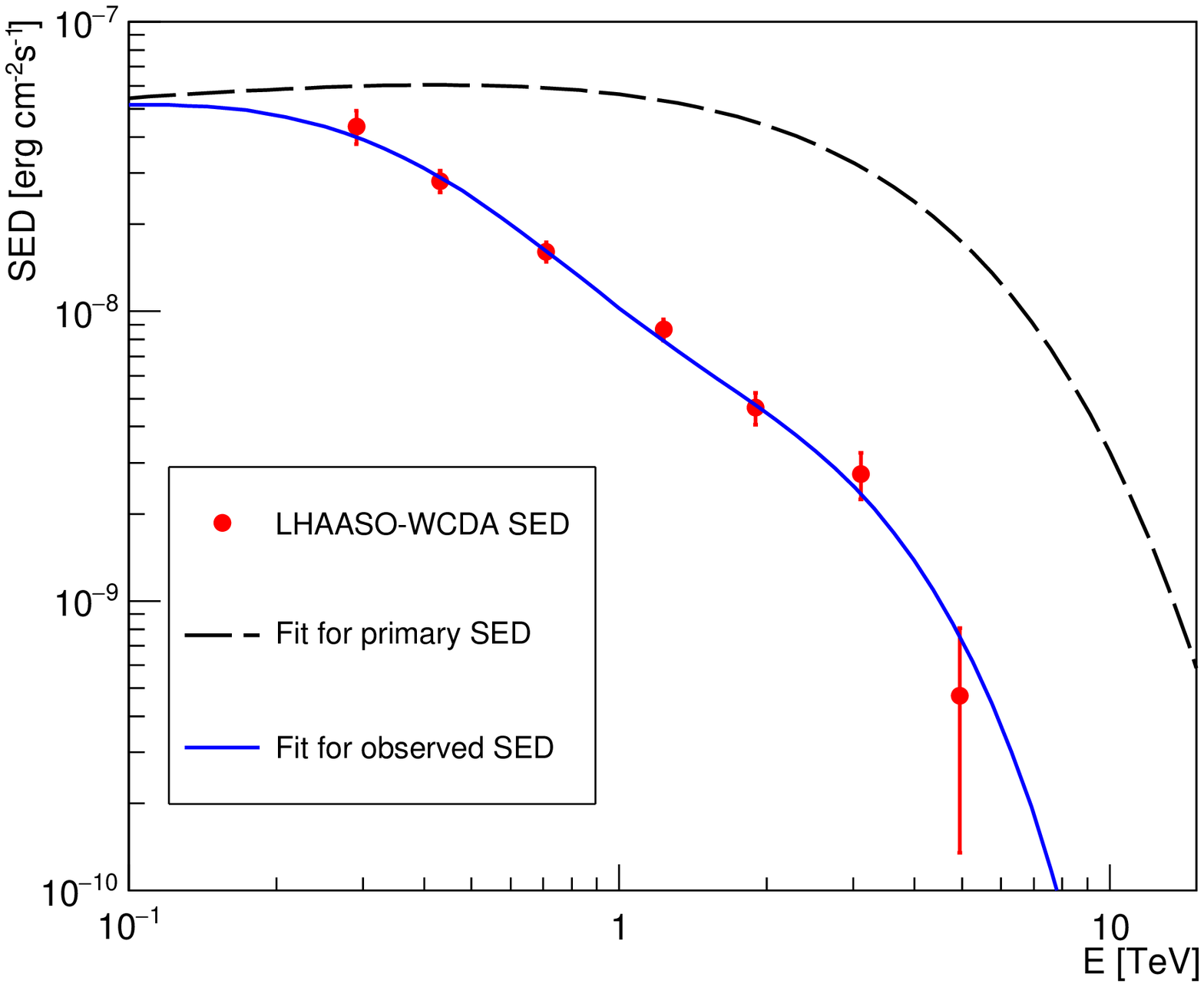}}\par
            \subcaptionbox{Squares: $230-300$~s, re-scaled to $231-326$~s, for which the curves are plotted\label{FigC6}}{\includegraphics[width=7.4cm]{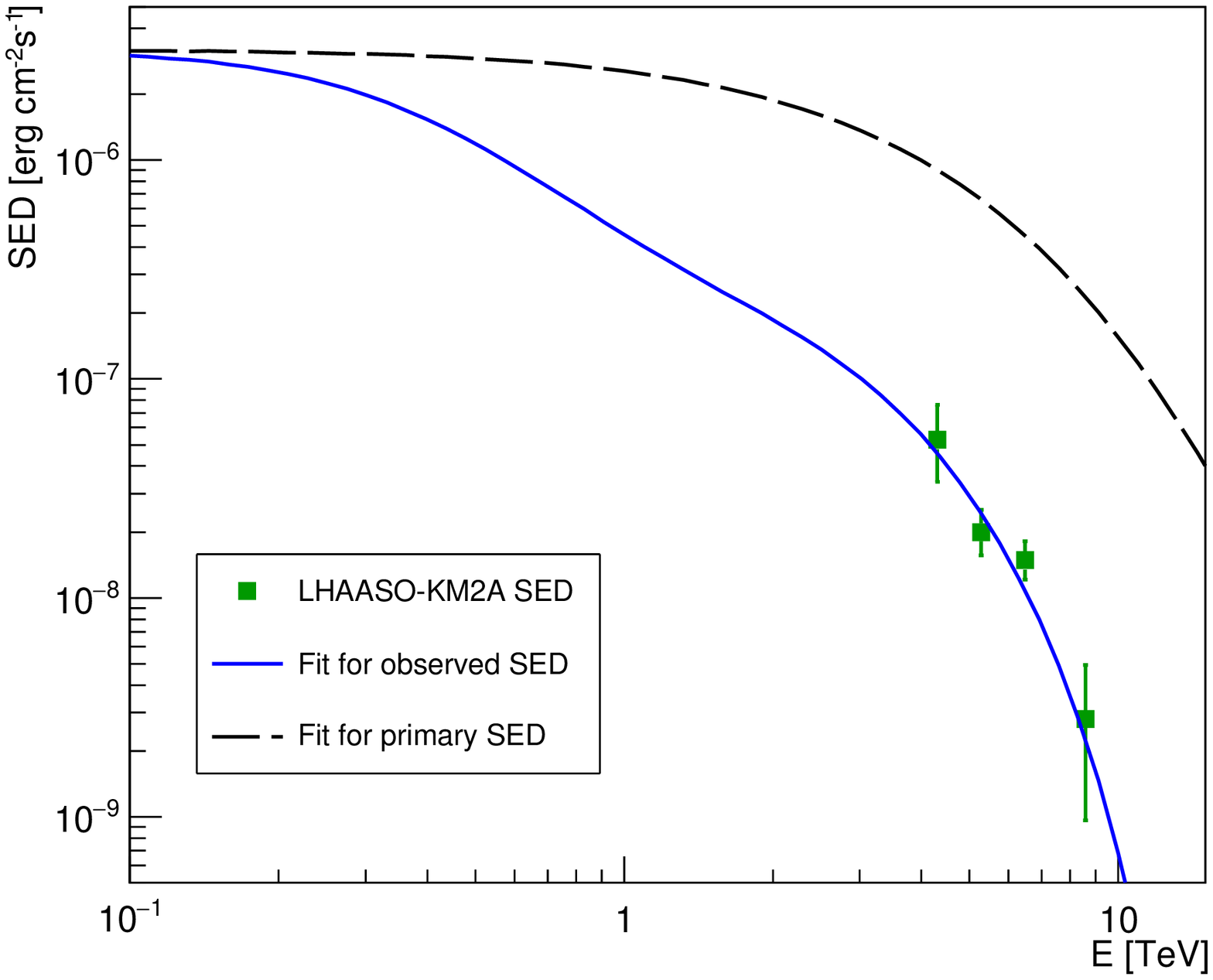}}\par
        \end{multicols}
\caption{SEDs observed by LHAASO-WCDA and LHAASO-KM2A in different time intervals after $T_0$ and the fitting curves to them. LHAASO-KM2A data points were rescaled to match the LHAASO-WCDA time intervals (see the text for more details.) \label{FigB}}
\end{figure*}

For the case of the fourth time interval (the 326-900 s time range) the dataset of LHAASO-KM2A \citep{Wang2023} for the time range 300--900~s was utilized as well. These time ranges are slightly different. To account for these conditions, a re-normalization of the LHAASO-KM2A SED was performed as follows. At first, the fluence correction factor was calculated:
\begin{equation}
K_{\mathrm{Fluence}} = \frac{\int\limits_{326\:s}^{900\:s} L(t)dt}{\int\limits_{300\:s}^{900\:s} L(t)dt} = 0.912,
\end{equation}
where $L(t)$ is the light curve. Eq.~(S10) from \citet{Cao2023} was assumed for the $\gamma$-ray light curve. Then the flux and the SED were calculated (see green squares in Fig~\ref{FigC4}). After that, the corrected LHAASO-KM2A SED was utilized in the fitting procedure in the same manner as the LHAASO-WCDA SED.

Finally, the LHAASO-KM2A dataset for the time range of 230--300 s is available \citep{Wang2023}. We perform the same correction procedure as introduced above for this dataset, re-scaling it to match the time range of 231--326 s (green squares in Fig~\ref{FigC6}). Then we compute the average SED over the time range of 231--326 s based on the LHAASO-WCDA dataset (blue curve in Fig~\ref{FigC6}). The LHAASO-KM2A SED is well fit by this curve; therefore, we do not introduce any further corrections to the fitting procedure concerning the 231--326 s time range.

\section{Cascade echo SEDs for a wide range of the magnetic field strength} \label{sec:appendixd}

\begin{figure}
\centering
\includegraphics[width=7.4cm]{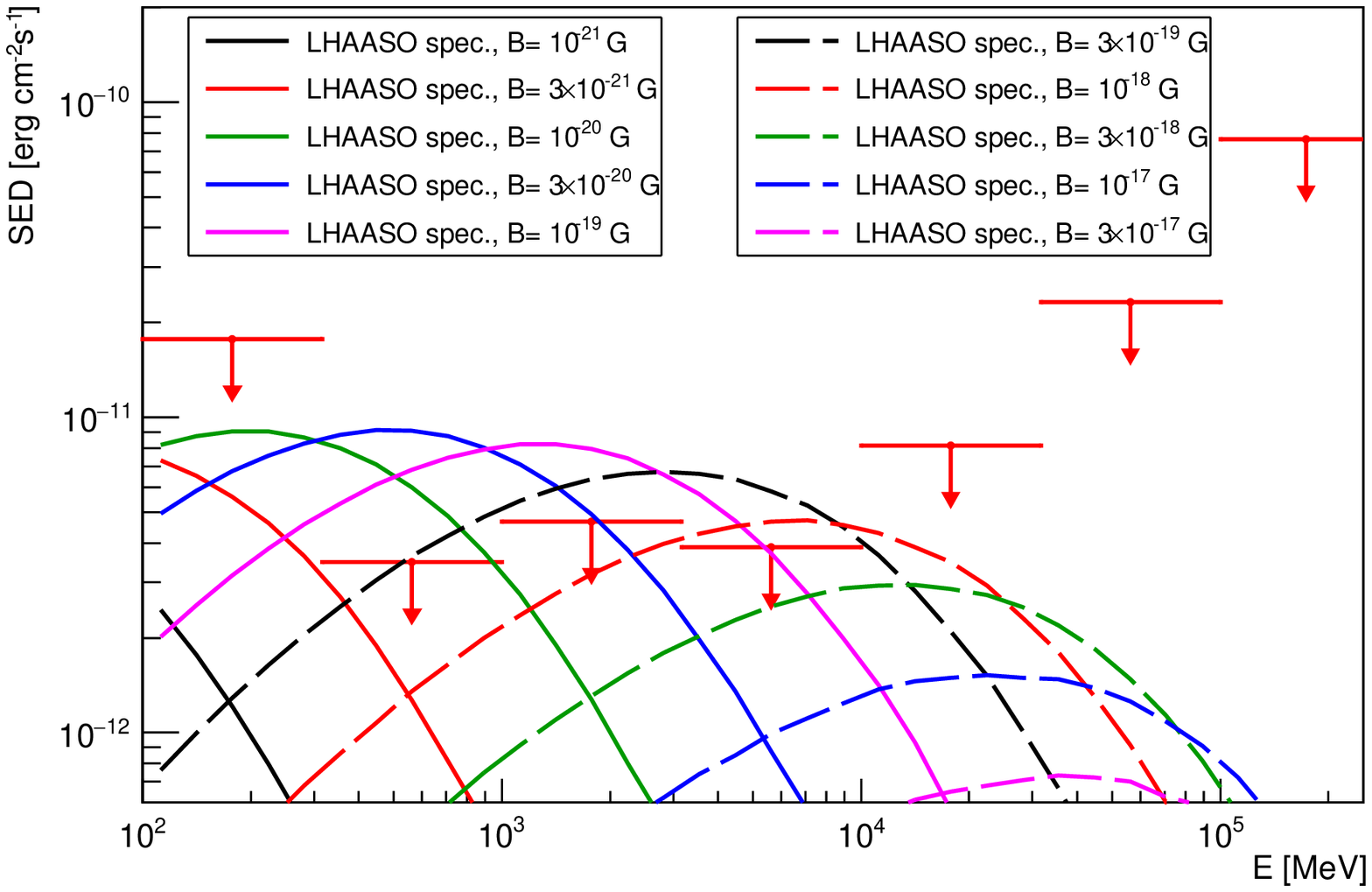}
\caption{Same as in Fig.~\ref{Fig5}, but for ten different values of $B$ (see the text for more details). In all cases $\delta T_{E} = 90$ days.}
\label{FigD1}
\end{figure}

In Fig.~\ref{FigD1} we show model curves for observable cascade SEDs for ten different values of $B$ for $\delta T_{E} = 90$ days. Solid curves denote $B = 10^{-21}$ G (black), $B = 3\times10^{-21}$ G (red), $B = 10^{-20}$ G (green), $B = 3\times10^{-20}$ G (blue), $B = 10^{-19}$ G (magenta). Dashed curves denote $B = 3\times10^{-19}$ G (black), $B = 10^{-18}$ G (red), $B = 3\times10^{-18}$ G (green), $B = 10^{-17}$ G (blue), $B = 3\times10^{-17}$ G (magenta).

\section{Comparison with the case of start time = 0} \label{sec:appendixe}

\begin{figure}
\centering
\includegraphics[width=7.4cm]{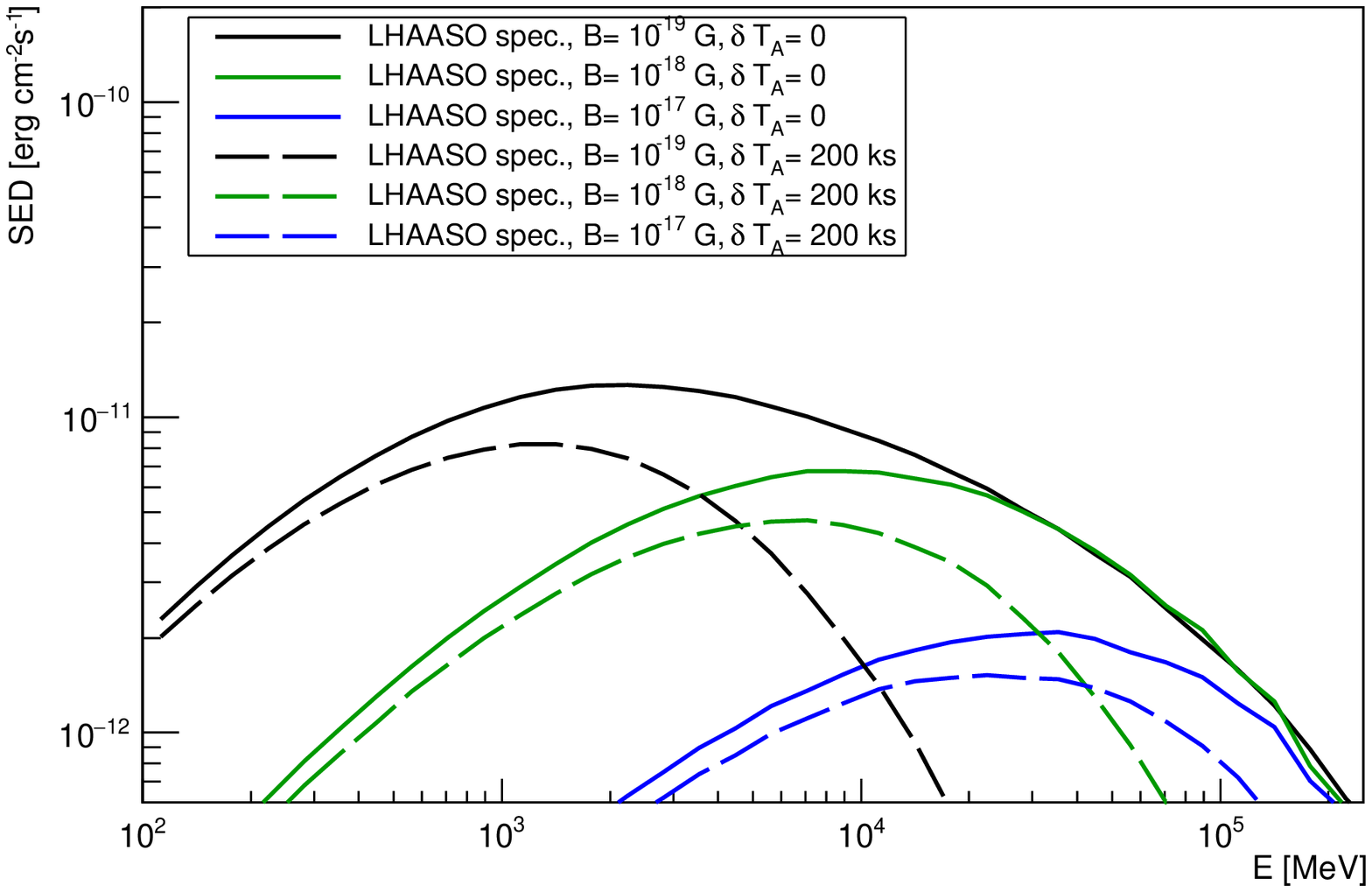}
\caption{Cascade echo SEDs for $\delta T_{E} = 90$ days and two values of $\delta T_A = 0$~s and $\delta T_A = 2 \times 10^{5}$~s.}
\label{FigE1}
\end{figure}

A comparison of the cascade echo SEDs for $\delta T_{E} = 90$ days and two values of $\delta T_A = 0$~s and $\delta T_A = 2 \times 10^{5}$~s is presented in Fig.~\ref{FigE1}. A part of high-energy cascade $\gamma$-rays arrive before $\delta T_A = 2 \cdot 10^{5}$~s and thus these $\gamma$-rays do not contribute to the model SEDs calculated for $\delta T_A = 2 \times 10^{5}$~s. This explains the high-energy cutoffs in the cascade echo SEDs.

\bsp	
\label{lastpage}
\end{document}